\documentstyle[preprint,aps,epsfig,eqsecnum]{revtex}

\newcommand{\Dst}{D^*}
\newcommand{\Dnot}{D_0}
\newcommand{\Done}{D_1}
\newcommand{\Dp}{D'^*}
\newcommand{\Bst}{B^*}
\newcommand{\Bnot}{B_0}
\newcommand{\Bone}{B_1}
\newcommand{\Bp}{B'^*}
\newcommand{\MB}{m_B}
\newcommand{\MD}{m_D}

\newcommand{\be}{\begin{eqnarray}}
\newcommand{\ee}{\end{eqnarray}}
\newcommand{\bl}{$B_{l4}$ }
\newcommand{\e}{\epsilon}
\renewcommand{\o}{\omega}
\newcommand{\eten}{\epsilon_{\mu\nu\alpha\beta}}
\renewcommand{\Re}{{\rm Re}}
\renewcommand{\Im}{{\rm Im}}
\def\sls#1{{#1 \!\!\! /}}

\tightenlines
\begin{document}

\newpage
\setcounter{page}{0}

\begin{titlepage}
 \begin{flushright}
 \hfill{hep-ph/9811395}\\
 \hfill{YUMS 98--019}\\
 \hfill{SNUTP 98--126}\\
 \hfill{KIAS-P98035}\\

 \end{flushright}
\vspace*{0.8cm}

\begin{center}
{\large\bf CP Violation in the Semileptonic $B_{l4}$ ($B \rightarrow D \pi l \nu$) 
 Decays: \\
A Model Independent Analysis}
\end{center}
\vskip 0.8cm
\begin{center}
{\sc C. S.~Kim$^{\mathrm{a,b,}}$\footnote{e-mail : kim@cskim.yonsei.ac.kr,
~~ http://phya.yonsei.ac.kr/\~{}cskim/},
 Jake Lee$^{\mathrm{a,}}$\footnote{e-mail : jilee@theory.yonsei.ac.kr} and 
 W.~Namgung$^{\mathrm{c,}}$\footnote{e-mail : ngw@cakra.dongguk.ac.kr}}

\vskip 0.5cm

\begin{small} 
$^{\mathrm{a}}$ Department of Physics, Yonsei University 120-749, Seoul, 
                  Korea \\
\vskip 0.2cm
$^{\mathrm{b}}$ School of Physics, Korea Institute for Advanced Study, 
            Seoul 130-012, Korea \\
\vskip 0.2cm
$^{\mathrm{c}}$ Department of Physics, Dongguk University 100-715, 
                  Seoul, Korea
\end{small}
\end{center}

\vspace{0.5cm}
\begin{center}
 (\today)
\end{center}

\setcounter{footnote}{0}
\begin{abstract}
CP violation from physics beyond the Standard Model is investigated in 
$B_{l4}$ decays: $B\to D\pi l\bar{\nu}_l$. 
The semileptonic $B$-meson decay to a $D$-meson 
with an emission of single pion is
analyzed with heavy quark effective theory and chiral perturbation theory.
In the decay process, we include various excited states as intermediate states 
decaying to the final hadrons, $D+\pi$. The CP violation is implemented 
in a model independent way, 
in which we extend leptonic current
by including complex couplings  of the scalar sector and those of the vector sector
in extensions of the Standard Model.
With these complex couplings,
we calculate the CP-odd rate asymmetry and the optimal asymmetry.
We find that the optimal asymmetry is sizable and
can be detected at $1\sigma$ level with about 
$10^6$-$10^7$ $B$-meson pairs, 
for some reference values of new physics effects.
\end{abstract}
\vskip 1cm
%
\end{titlepage}

\newpage
\baselineskip .29in
\renewcommand{\thefootnote}{\alph{footnote}}

\section{Introduction}

\noindent
Semileptonic 4-body decays of mesons, such as $K_{l4}$, $D_{l4}$ and $B_{l4}$
with emission of a pion have been studied in detail by many authors  
\cite{KL,DL,BL,goity}.
$B_{l4}$ decays will be fully investigated soon at the forthcoming 
$B$-meson factories.
The decays of $K_{l4}$, $D_{l4}$, and $B_{l4}$ have distinct characteristics 
in many theoretical aspects:
$K_{l4}$ decays can be studied with chiral perturbation theory  
for full decay phase space \cite{KL}.
However, in $D_{l4}$ decays the chiral perturbation theory can be applied only 
to a very small region of 
the final state phase space  \cite{DL} since light mesons in final state have
relatively large energies.
The pions emitted in $B_{l4}$ decays have such a wide momentum range that
one may have difficulties to analyze the decays over whole phase space.
However, if we restrict our attention to the soft pion limit,
we could investigate $B_{l4}$ decays, including the final state $D$-meson, 
by the combined method of heavy quark expansion and chiral 
perturbation expansion \cite{BL,goity}.
The significance of $B\to D\pi l\nu$ decay mode is seen from the observation that
the elastic modes of $B\to De\nu$ and $B\to D^*e\nu$ account 
for less than $70\%$ of the
total semileptonic branching fraction.
Goity and Roberts \cite{goity} have studied $B\to D\pi l\nu$ decays by including
various intermediate states which are decaying to $D + \pi$,
and found that the effects of higher excited intermediate
states are substantial compared to the lowest state  of $D^*$, 
as the invariant mass of 
$D\!+\!\pi$ grows away from the ground $D^*$ resonance region.

In this work, we consider the possibility of probing direct CP violation 
in the decay of $B^\pm \to D \pi l^\pm \nu$  in a model independent way,
in which we extend leptonic current
by including complex couplings  of the scalar sector and those of the vector sector
in extensions of the Standard Model (SM).
In our separate paper \cite{bcp}, we also consider the phenomenon
within specific models such as 
multi-Higgs-doublet model
and scalar leptoquark models.
In order to observe direct CP violation effects, there should exist interferences 
not only 
through weak CP-violating
phases but also different CP-conserving strong phases.
In \bl decays, CP-violating phases can be generated through interference
between $W$-exchange diagrams and 
scalar-exchange diagrams with complex couplings. 
The CP-conserving phases may come from the absorptive parts of 
the intermediate resonances in \bl decays. 

CP violation can be also investigated through  T-odd momentum
or spin correlations in semileptonic heavy meson decays to three or more final state 
particles in some extended models \cite{garisto,korner}.
Especially, the authors of Ref.~ \cite{korner}
analyzed the possibility of probing CP-violation by extracting T-odd angular 
correlations in the lowest resonance decay of $B\to D^*(\to D\pi)l\nu$ 
and found that the effects can be detected in some cases.
Here we include higher excited states, such as $P$-wave, $D$-wave states 
and the first radially excited $S$-wave
state, as intermediate states in $B\to D\pi l\nu$ decay, since these
intermediate states could contribute substantially to the decay \bl \cite{goity}.
As can be seen later, including these higher excited states would significantly
amplify the CP violation effects.

In previous studies of $B_{l4}$, decays to only light leptons ($e$ and $\mu$) have been 
considered,
because the width is very small in the case of heavy leptonic decay, $B_{\tau 4}$.
However, the CP violation effects implemented by interference 
between vector boson and
scalar boson exchanges in leptonic current may in general be proportional to 
the lepton's mass.
Therefore, we here consider the case including a tau lepton as well.
We note  the larger lepton mass implies the smaller available region of hadronic 
invariant mass in the final state phase space, which actually means 
the relatively larger region of phase space to which chiral perturbation theory 
is applicable 
in $B_{\tau 4}$ decays, compared to light leptonic decay cases.

In Section II, we briefly review the heavy quark effective theory 
and chiral perturbation
theory to calculate the amplitude of $B\to D\pi l\nu$ decay.
The detailed formalism dealing with \bl decay is given in Section III and Appendices. 
And the observable asymmetries are shown in Section IV. 
Section V contains our results, discussions and conclusions.

\section{Heavy Quark Effective Theory and Chiral Perturbation Theory}

\noindent
The formalism of \bl decay in the context of heavy quark effective theory (HQET) and 
chiral perturbation theory (chPT) has been studied by many authors \cite{BL,goity}.
We follow
the procedure described in that literature, and make it suitable for our
analysis by including various intermediate states and retaining lepton masses.
We briefly review these theories and describe explicitly 
our procedure of obtaining the amplitude of the decay.

The interactions of the octet of pseudogoldstone (pG) bosons 
with hadrons containing a single
heavy quark are constrained by two independent symmetries: 
spontaneously broken chiral
SU(3)$_L\times $ SU(3)$_R$ symmetry and heavy quark spin-flavor SU$(2N_h)$ 
symmetry \cite{HQET},
where $N_h$ is the number of heavy quark flavors.
Within the frame work of HQET \cite{spect},
the spin $J$ of a meson consists of the spin of a heavy quark ($Q$),
the spin of a light quark ($q$) and relative angular momentum $l$:
\be
J=S_Q+S_l+l\;.
\ee
We denote the meson state as $J^P$ with its parity.
If we define $j=S_l+l$ which corresponds to the spin of the light component of 
the meson, we have a multiplet for each $j$:
\be
J=j\pm \frac{1}{2}\;.
\ee
Then, for a meson $M$, HQET predicts the following multiplets up to $l=2$:
\be
l=0:&& (0^-,1^-)\qquad (M,M^*)\;,\nonumber\\
l=1:&& (0^+,1^+)\qquad (M_0,M_1^{(0)})\nonumber\\
    && (1^+,2^+)\qquad (M_1^{(1)},M_2)\;,\nonumber\\
l=2:&& (1^-,2^-)\qquad (M_1,M_2^{(0)})\nonumber\\
    && (2^-,3^-)\qquad (M_2^{(1)},M_3)\;,
\ee
where we denote the corresponding meson states by the notation in the last column.
Furthermore, there could be radially excited states.
For example, the first radially excited states are
\be
l=0\;(n=2):\; (0^-,1^-)\qquad (M',M'^*)\;.
\ee
Among these resonances (denoted by $\tilde{M}$), $\tilde{D}\to D\pi$ or
$B\to \tilde{B}\pi$ decay is possible only for $J^P=0^+$, $1^-$, $2^+$ and $3^-$
resonances because of parity conservation.
However, if we use chiral expansion, the decay amplitude of $2^+$ state, $M_2$, is 
proportional to $(p_\pi)^2$, and 
that of $3^-$ state is proportional to $(p_\pi)^3$  \cite{goity}, 
so their contributions will be suppressed in the soft pion limit.
Therefore, in the leading order in $p_\pi$, the resonances contributing
to $B_{l4}$ decays are
\be
\tilde{M}=M^*,\; M_0,\; M_1\;{\rm and}\; M'^*,
\ee
where $M$ stands for $B$ or $D$ meson.

The weak decay matrix elements of $B$ to $D$ are given in the heavy quark 
limit \cite{goity} as
\be
\langle \Dst(v',\e)|H_\mu|B(v)\rangle &=&\xi(\o)[-i\eten v'^\alpha v^\beta 
           +g_{\mu\nu}(\o+1)-v'_\mu  v_\nu]\e^{*\nu},\nonumber\\
\langle \Dnot(v')|H_\mu|B(v)\rangle &=&-\rho_1(\o)(v-v')_\mu,\nonumber\\
\langle \Done(v',\e)|H_\mu|B(v)\rangle 
  &=&\frac{1}{\sqrt{6}}\rho_2(\o)[i\eten v^\alpha 
          v'^\beta(\o-1)+g_{\mu\nu}(\o^2-1)\nonumber\\
          &&-v_\nu\{(2+\o)v'_\mu-3v_\mu\}]
          \e^{*\nu},\nonumber\\
\langle \Dp(v',\e)|H_\mu|B(v)\rangle &=&\xi^{(1)}(\o)[-i\eten v'^\alpha v^\beta
          +g_{\mu\nu}(\o+1)-v'_\mu v_\nu]\e^{*\nu},
\label{vforms}
\ee 
where $H_\mu = \bar{c}\gamma_\mu(1-\gamma_5)b=V_\mu-A_\mu$ and $\o \equiv v\cdot v'$.
The effective currents where a $B$-meson resonance decays into a ground-state
$D$-meson are easily obtained by simply taking the hermitian conjugate of the
currents given above, followed by interchanging the symbols 
$B\leftrightarrow D$ and $v\leftrightarrow v'$.

For the form factors $\xi(\o),\xi^{(1)}(\o),\rho_1(\o)$ and $\rho_2(\o)$ 
in Eq.~(\ref{vforms}),
we adopt the forms of Ref.~ \cite{goity}, which were derived from the quark model.
The explicit forms are
\be
\xi(\o)&=&\exp\left[\frac{\overline{\Lambda}^2}{4\beta^2}(\o^2-1)\right],\nonumber\\
\xi^{(1)}(\o)&=&-\sqrt{\frac{2}{3}}
     \left[\frac{\overline{\Lambda}^2}{4\beta^2}(\o^2-1)\right]
     \exp\left[\frac{\overline{\Lambda}^2}{4\beta^2}(\o^2-1)\right],\nonumber\\
\rho_1(\o)&=& \frac{1}{\sqrt{2}}\frac{\overline{\Lambda}}{\beta}
              \left(\frac{2\beta\beta'_1}{\beta^2+\beta^{\prime 2}_1}\right)^{5/2}
              \exp\left[\frac{\overline{\Lambda}^2}{2(\beta^2+\beta^{\prime 2}_1)}
                 (\o^2-1)\right],\nonumber\\
\rho_2(\o)&=& \frac{1}{2\sqrt{2}}\left(\frac{\overline{\Lambda}}{\beta}\right)^2
              \left(\frac{2\beta\beta'_2}{\beta^2+\beta^{\prime 2}_2}\right)^{7/2}
              \exp\left[\frac{\overline{\Lambda}^2}{2(\beta^2+\beta^{\prime 2}_2)}
                 (\o^2-1)\right],
\label{formfactor}
\ee
where $\overline{\Lambda}$ is defined by writing the mass of the ground state
as $M_{(0^-,1^-)}=m_Q+\overline{\Lambda}$ with $m_c=1628$ MeV and 
$m_b=4977$ MeV \cite{Godfrey}, 
and numerical values are 
$\beta=0.29,\; \beta'_1=0.28$ and $\beta'_2=0.26$.

In order to deal with the interactions of pG bosons with hadrons containing
a single heavy quark, one may construct an effective lagrangian with the
two symmetries above and perform a simultaneous expansion in the momenta of pG
bosons and the inverse masses of the heavy quarks.
Such a lagrangian has been described in Ref.~ \cite{elag}
for heavy hadrons with the light degrees of freedom in the ground state.
For construction of such effective lagrangians, it is convenient and now common
to introduce superfields associated with each multiplet \cite{superm}.
Using the superfield formalism, one may represent the spin-flavor 
symmetry explicitly, and
according to a velocity superselection rule \cite{sselect}, one can have one such
superfield assigned with each four velocity $v_\mu$ at the leading order
in the inverse heavy mass expansion.

The superfield for the ground-state heavy meson multiplet $(0^-,\;1^-)$ with velocity
$v_\mu$ is 
\be
{\cal H}_-=\frac{1+\sls{v}}{2}(-\gamma_5 P+\gamma^\mu V^*_\mu),
\ee
where $P$ and $V^*_\mu$ are the fields associated with the pseudoscalar and vector
partners, respectively.
This multiplet transforms under spin-symmetry operations as
\be
&& {\cal H}_-\to \exp(-i\vec{\epsilon}\cdot \vec{S}_v){\cal H}_-,\nonumber\\
&& {\rm with}~~~S_v^j=i\epsilon^{jkl}[{\sls e}_k,{\sls e}_l]\frac{(1+{\sls v})}{2},
\ee
where $e^\mu_k,\; k=1,2,3$, are space-like vectors orthogonal to 
the four-velocity $v_\mu$.
Similarly, one can associate superfields with excited states \cite{exsuper}.
The multiplets of our interest, $(0^+,1^+),\; (1^-,2^-)$ and $(0^-,1^-)'$ 
are described
by the superfields:
\be
&&{\cal H}_+=\frac{1+\sls{v}}{2}\left(-H_{0^+}
   +\gamma^\mu\gamma_5 H^\mu_{1^+}\right),\nonumber\\
&&{\cal H}_-^\mu=\frac{1+\sls{v}}{2}\left(\sqrt{\frac{3}{2}}H^\nu_{1^-}[g^\mu_\nu
               -\frac{1}{3}\gamma_\nu(\gamma^\mu+v^\mu)]
              +\gamma_5\gamma_\nu H^{\mu\nu}_{2^-}\right),\nonumber\\
&&{\cal H}'_-=\frac{1+\sls{v}}{2}\left(-\gamma_5H'_{0^-}
+\gamma_\mu H'^\mu_{1^-}\right),
\ee
respectively. All the tensors are traceless, symmetric and 
transverse to the four-velocity.
These excited superfields transform under the spin-symmetry operations 
in the same way 
as the ground state superfield does.

The strong interactions of heavy mesons with pG bosons are described by
the so-called ``heavy-light'' chiral lagrangian, which is written 
in terms of the usual
exponentiated matrix of pG bosons,
\be
\Sigma=\exp\left(\frac{2i{\cal M}}{\sqrt{2}f_\pi }\right),
\ee
where ${\cal M}$ is a $3\times 3$ matrix for the octet of pG bosons,
\be
{\cal M}=\left(\begin{array}{ccc}
   \frac{\pi^0}{\sqrt{2}}+\frac{\eta}{\sqrt{6}}&\pi^+&K^+\\
   \pi^-&-\frac{\pi^0}{\sqrt{2}}+\frac{\eta}{\sqrt{6}}&K^0\\
   K^-&\overline{K}^0&-\sqrt{\frac{2}{3}}\eta 
   \end{array}\right),
\ee
and
\be
f_\pi  =93\;{\rm MeV}
\ee
is the pion decay constant. The lagrangian for the pG bosons is
\be
{\cal L}_{\cal M}=\frac{f_\pi ^2}{4}{\rm tr}\partial_\mu\Sigma^\dagger 
\partial^\mu\Sigma,
\ee
which contains all SU$(3)_L\times$ SU$(3)_R$ invariant interactions up to two
derivatives among the pG bosons. One can easily see the invariance of the lagrangian
${\cal L}_{\cal M}$ under SU$(3)_L\times$ SU$(3)_R$ chiral transformation, 
since in this representation the unitary matrix transforms as
\be
\Sigma\to \Sigma'=L\Sigma R^\dagger,
\ee
where $L$ and $R$ are global transformations in SU$(3)_L$ and SU$(3)_R$, 
respectively.
Since the chiral symmetry SU$(3)_L\times$ SU$(3)_R$ is spontaneously broken down to
SU$(3)_V$, we can deal with interactions of such pG bosons to the usual hadrons 
by introducing a new matrix \cite{gbook}
\be
\xi=\Sigma^{1/2},
\ee
which transforms under the SU$(3)_L\times$ SU$(3)_R$ as
\be
\xi\to \xi'=L\xi U^\dagger =U\xi R^\dagger,
\ee
where $U$ is a unitary matrix depending on $L$ and $R$ and nonlinear functions of
pG fields. From $\xi$ we can construct a vector field $V_\mu$ and
an axial vector field $A_\mu$ as follows:
\be
&&V_\mu=\frac{1}{2}(\xi^\dagger\partial_\mu\xi+\xi\partial_\mu\xi^\dagger)
       \simeq\frac{1}{4f_\pi ^2}[{\cal M},\;\partial_\mu{\cal M}],\\
&&A_\mu=\frac{i}{2}(\xi^\dagger\partial_\mu\xi-\xi\partial_\mu\xi^\dagger)
       \simeq\frac{-1}{\sqrt{2}f_\pi }\partial_\mu{\cal M}.
\ee
The vector field transforms like a gauge field under chiral transformation
\be
V_\mu\to V'_\mu=UV_\mu U^\dagger+U\partial_\mu U^\dagger,
\ee
and the axial vector field is an SU$(3)$ octet which transforms as
\be
A_\mu\to A'_\mu=UA_\mu U^\dagger.
\ee
The heavy quark spin-symmetry multiplet, ${\cal H}$, transforms under
 a chiral rotation as
\be
{\cal H}\to U{\cal H}.
\ee
Note that all multiplets we are considering are isotriplets, 
and effectively reduce to isodoublets
since we do not include the strange quark in our consideration.

In order to construct chirally invariant lagrangian, 
we make use of a gauge-covariant
derivative involving $V_\mu$:
\be
\nabla_\mu = \partial_\mu+V_\mu.
\ee
Using this covariant derivative and the axial vector field $A_\mu$, one can construct
an effective lagrangian, which possesses spin-flavor and chiral symmetry explicitly.
In Ref.~\cite{goity}, the strong interaction effective lagrangian is given 
to the lowest chiral order, {\it i.e.\/}, 
to $O(p_\pi)$ in the heavy meson sector and to $O(p_\pi^2)$ in the pion sector.
Among many possible terms in the lagrangian, the part of the interaction between 
heavy mesons and pions is
\be
{\cal L}_{int}&=& g{\rm tr}(\overline{{\cal H}}_-A^\mu{\cal H}_-\gamma_\mu\gamma_5)
     +\alpha_1{\rm tr}(\overline{{\cal H}}_+A^\mu{\cal H}_-\gamma_\mu\gamma_5)\nonumber\\
        &&+\alpha_2{\rm tr}(\overline{{\cal H}}_{-\mu}A^\mu{\cal H}_-\gamma_5)
         +\alpha_3{\rm tr}(\overline{{\cal H}}'_-A^\mu{\cal H}_-\gamma_\mu\gamma_5),
\ee
where $\overline{\cal H}=\gamma^0 {\cal H}^\dagger \gamma^0$, 
and $g$ and $\alpha_i$'s are the low energy coupling constants.
From the above lagrangian, one can obtain vertices for soft neutral pion 
emission \cite{goity}:
\be
(DD^{*\mu})\pi^0 &:& \frac{g}{f_\pi }p_\pi^\mu,\nonumber\\
(DD^{0^+})\pi^0 &:& \frac{\alpha_1}{f_\pi }p_\pi\cdot v,\nonumber\\
(DD^{\mu}_{1^-})\pi^0 &:& 
\sqrt{\frac{2}{3}}\frac{\alpha_2}{f_\pi }p_\pi^\mu,\nonumber\\
(DD'^{*\mu})\pi^0 &:& \frac{\alpha_3}{f_\pi }p_\pi^\mu.
\ee
Using these vertex factors and the heavy meson transition matrix elements 
in Eq.~(\ref{vforms}),
one can calculate $B_{l4}$ decay amplitudes.

\section{\bl Decay Rates with Scalar Couplings}

\noindent
We consider $B_{l4}$ decays of $B^-\to D^+\pi^-l\bar{\nu}$ and
$B^-\to D^0\pi^0l\bar{\nu}$.
We here show theoretical expressions for $B^-\to D^0\pi^0l\bar{\nu}$.
For $B^-\to D^+\pi^-l\bar{\nu}$, we can easily derive them from those for 
$B^-\to D^0\pi^0l\bar{\nu}$ at the amplitude level by using isospin relation.
The amplitude  has the general form:
\be
T=\kappa \Big[(1+\chi)j_\mu\Omega_V^\mu+\eta j_s \Omega_s\Big],\qquad 
\kappa=V_{cb}\frac{G_F}{\sqrt{2}}\sqrt{\MB\MD},
\ee
where $j_\mu$ is the $(V-A)$ charged leptonic current and $j_s$ is 
a Yukawa-type scalar current in the leptonic sector.
Here the parameters $\chi$ and $\eta$, which parametrize contributions from physics
beyond the SM, are in general complex. Note that the SM values are
$\chi=\eta =0$.
We retain charged lepton mass since we also consider decays of $B_{\tau 4}$.
The vector interaction part of the hadronic amplitude, $\Omega_V^\mu$, 
receives contributions from two types of
diagrams, illustrated in Fig.~1(a) and 1(b), respectively:
\be
\Omega_V^\mu&=&\langle D\pi|\bar{c}\gamma^\mu(1-\gamma_5)b|B\rangle \nonumber\\
    &=&\sum_{\tilde{B}_i}\langle D|\bar{c}\gamma^\mu(1-\gamma_5)b|\tilde{B}_i\rangle 
          \langle \tilde{B}_i\pi|B\rangle 
          +\sum_{\tilde{D}_i}\langle D\pi|\tilde{D}_i\rangle \langle \tilde{D}_i|
          \bar{c}\gamma^\mu(1-\gamma_5)b|B\rangle ,
\ee
and $\Omega_s$ is the corresponding scalar current matrix element:
\be
\Omega_s=\langle D\pi|\bar{c}(1-\gamma_5)b|B\rangle ,
\ee
where $\tilde{D}_i$ and $\tilde{B}_i$ stand for intermediate excited states of
our interest.
We can obtain Yukawa interaction form factors by multiplying the $(V-A)$
currents with momentum transfer $q^\mu$:
\be
q^\mu = p_B-p_D=\MB v^\mu - \MD v'^\mu,
\ee
\be
&&\langle \Dst(v',\e)|q^\mu A_\mu|B(v)\rangle 
=-\xi(\o)(\MB+\MD)(v\cdot\e^*),\nonumber\\
&&\langle \Dnot(v')|q^\mu A_\mu|B(v)\rangle =\rho_1(\o)(\MB+\MD)(1-\o),\nonumber\\
&&\langle \Done(v',\e)|q^\mu A_\mu|B(v)\rangle 
=-\frac{2}{\sqrt{6}}\rho_2(\o)(\MB+\MD)(1-\o)
                     (v\cdot\e^*),\nonumber\\
&&\langle \Dp(v',\e)|q^\mu A_\mu|B(v)\rangle =-\xi^{(1)}(\o)(\MB+\MD)(v\cdot\e^*);\\
&&{ }\nonumber\\
&&\langle {\rm all~ states} |q^\mu V_\mu|B(v)\rangle =0 .
\label{sforms}
\ee
And using 
\be
q^\mu V_\mu=(\MB-\MD)\bar{c}b,\qquad  q^\mu A_\mu=-(\MB+\MD)\bar{c}\gamma_5 b,
\label{qva}
\ee
in the heavy quark limit, one can relate any Yukawa-type interaction
with the vector (axial) current ones.
Consequently, we get the following relation
\be
q_\mu \Omega_V^\mu=(m_B+m_D)\Omega_s.
\ee
We define the dimensionless parameter $\zeta$, 
which determines the size of the 
scalar contributions relative to the vector ones:
\be
\zeta=\frac{\eta}{1+\chi}.
\label{zeta}
\ee
Note that the SM corresponds to the case with $\chi=\eta =\zeta =0$.
Then the amplitude can be written as
\be
T=\kappa(1+\chi)j_\mu\Omega^\mu,\qquad \Omega^\mu
        =\Omega_V^\mu+\zeta\Omega_s\frac{L^\mu}{m_l},
\ee
where we used the Dirac equation for leptonic current,
$L^\mu j_\mu =m_l j_s$ with $L^\mu=p_l^\mu+p_{\bar{\nu}}^\mu$.

Then, the relevant Feynman diagrams read
\be
\Omega_\mu&=&
   \frac{g}{f_\pi }\xi(\o)p_\nu \bigg\{ \Theta^{\nu\rho}(v)\Delta(\Bst)
   \Big[iv^\alpha v'^\beta\e_{\mu\alpha\beta\rho}+g_{\mu\rho}(1+\o)
   -v_\mu v'_\rho-\zeta v'_\rho \frac{L_\mu}{m_l}\Big]\nonumber\\
&& +\Theta^{\nu\rho}(v')\Delta(\Dst)
   \Big[iv^\alpha v'^\beta\e_{\mu\alpha\beta\rho}+g_{\mu\rho}(1+\o)
   -v_\rho v'_\mu+\zeta v_\rho \frac{L_\mu}{m_l}\Big] \bigg\}\nonumber\\
&& +\frac{\alpha_1}{f_\pi }\rho_1(\o)\{(v-v')_\mu+\zeta(1-\o)\frac{L_\mu}{m_l}\}
   \Big[-\Delta(\Bnot)p\cdot v+\Delta(\Dnot)p\cdot v'\Big]\nonumber\\
&& +\frac{\alpha_2}{3f_\pi }\rho_2(\o)p_\rho \bigg\{ \Theta^{\nu\rho}(v)\Delta(\Bone)
   \Big[i\e_{\mu\nu\alpha\beta}v^\alpha v'^\beta (\o-1)
      +g_{_\mu\nu}(\o^2-1)\nonumber\\
&& -v'_\nu(v_\mu(2+\o)-3v'_\mu)+2\zeta(1-\o)v'_\nu \frac{L_\mu}{m_l}\Big]
    +\Theta^{\nu\rho}(v')\Delta(\Done)
   \Big[i\e_{\mu\nu\alpha\beta}v^\alpha v'^\beta (\o-1)\nonumber\\
&&  +g_{_\mu\nu}(\o^2-1)
-v_\nu(v'_\mu(2+\o)-3v_\mu)+2\zeta(1-\o)v_\nu \frac{L_\mu}{m_l}\Big] \bigg\}\nonumber\\
&&  \frac{\alpha_3}{f_\pi }\xi^{(1)}(\o)p_\nu \bigg\{ \Theta^{\nu\rho}(v)\Delta(\Bp)
   \Big[iv^\alpha v'^\beta\e_{\mu\alpha\beta\rho}+g_{\mu\rho}(1+\o)
   -v_\mu v'_\rho-\zeta v'_\rho \frac{L_\mu}{m_l}\Big]\nonumber\\
&& +\Theta^{\nu\rho}(v')\Delta(\Dp)
   \Big[iv^\alpha v'^\beta\e_{\mu\alpha\beta\rho}+g_{\mu\rho}(1+\o)
   -v_\rho v'_\mu+\zeta v_\rho \frac{L_\mu}{m_l}\Big]\bigg\},
\label{Rho}
\ee     
where $\Theta^{\mu\nu}(v)=g^{\mu\nu}-v^\mu v^\nu$, and
for numerical values of the coupling constants we use the predictions 
of Ref.~\cite{goity}:
\be
g=0.5,\quad \alpha_1=-1.43,\quad \alpha_2=-0.14,\quad \alpha_3=0.69.
\ee
Here the functions representing the propagator of the intermediate resonance,
$\Delta(\tilde{M})$, are defined as follows:
\be
\Delta(\tilde{B})&=&\frac{M_{\tilde{B}}}{(p_B-p)^2-M^2_{\tilde{B}}
       +iM_{\tilde{B}}\Gamma_{\tilde{B}}}
      =\frac{1}{-2(v\cdot p+\delta m_{\tilde{B}})+i\Gamma_{\tilde{B}}},\nonumber\\
\Delta(\tilde{D})&=&\frac{M_{\tilde{D}}}{(p_D+p)^2-M^2_{\tilde{D}}
      +iM_{\tilde{D}}\Gamma_{\tilde{D}}}
     =\frac{1}{2(v'\cdot p-\delta m_{\tilde{D}})+i\Gamma_{\tilde{D}}},
\ee
where we have incorporated the finite width of the resonances, 
$\Gamma_{\tilde{D}}$ and $\Gamma_{\tilde{B}}$.
This inclusion of resonance widths corresponds to considering the overall strong
interaction contribution, and plays an important role in our investigation.
Those widths serve as CP-conserving phases needed for direct CP violations.
Here the mass differences between the resonances and the ground-state mesons,
$\delta m_{\tilde{M}}$, are defined as $\delta m_{\tilde{M}}=m_{\tilde{M}}-m_M$,
and $\Gamma_{\tilde{M}}$ is the total width of the resonance.
For the specific values of the above quantities, we adopt the predictions 
in Ref.~\cite{goity}:
\be
&& \delta m_{M_0}=0.39\;{\rm GeV},\qquad \Gamma_{M_0}=1040\;{\rm MeV},\nonumber\\
&& \delta m_{M_1}=0.71\;{\rm GeV},\qquad 
   \Gamma_{M_1}=405\;{\rm MeV},\nonumber\\
&& \delta m_{M^{\prime *}}=0.56\;{\rm GeV},\qquad 
   \Gamma_{M^{\prime *}}=191\;{\rm MeV},
\ee
where $M$ stands for $B$ or $D$.
Note that we used the same numerical values for $\tilde{B}$ and $\tilde{D}$,
for simplicity, following Ref.~\cite{goity}. For $B_{l4}$ decays,
values for $\tilde{D}$ are much more important, and are also more precisely known 
from experiments compared to  values of $\tilde{B}$.
In order to do detailed calculations, first, we now define the kinematic variables
\be
&&P=p_D+p_\pi,~~\;~~{\rm with}~~p_D=m_Dv',~~\; p_\pi=p,\nonumber\\
&&Q=p_D-p,\nonumber\\
&&L=p_l+p_{\bar{\nu}},\nonumber\\
&&N=p_l-p_{\bar{\nu}}.
\ee
In terms of these variables, the most general form of $\Omega_\mu$ can be written as
\be
\Omega_\mu=\frac{i}{2}H\e_{\mu\nu\rho\sigma}L^\nu Q^\rho P^\sigma
          +FP_\mu+GQ_\mu +RL_\mu,
\label{Rho2}
\ee
where $H$, $F$, $G$ and $R$ are the form factors depending on three invariants
$\o$, $p\cdot v$ and $p\cdot v'$.
These form factors are easily obtained from the expressions in Eq.~(\ref{Rho}),
and their explicit forms are given in Appendix A.
We get the following invariants:
\be
P^2&=&s_M,\nonumber\\
Q^2&=&2(\MD^2+m_\pi^2)-s_M,\nonumber\\
L^2&=&s_L,\nonumber\\
N^2&=&-s_L+2m_l^2,\nonumber\\
P\cdot Q&=&\MD^2-m_\pi^2,\nonumber\\
P\cdot L&=& \frac{1}{2}(\MB^2-s_L-s_M),\nonumber\\
L\cdot N&=&m_l^2.
\ee

It is well known that there are five independent
kinematic variables for these processes 
when the spins of the initial and final states are zero or not observed.
From the momenta of the $B$-meson,
$D$-meson, the pion, the lepton, and its neutrino $p_B$, $p_D$, $p$, $p_l$, 
and $p_{\bar{\nu}}$, for the five independent variables we choose 
$s_M=(p_D+p)^2$, $s_L=(p_l+p_{\bar{\nu}})^2$,
$\theta$ ({\it i.e.\/}, the angle between the $D$ momentum in the $D\pi$ rest frame
and the moving direction of the $D\pi$ system in the $B$-meson's rest frame), 
$\theta_l$ ({\it i.e.\/}, the angle between the lepton momentum in the $l\bar{\nu}$
rest frame and the moving direction of the $l\bar{\nu}$ system in the
$B$-meson's rest frame) and $\phi$ ({\it i.e.\/}, the angle between the two decay planes
defined by the pairs (${\bf p},{\bf p}_D$) and 
(${\bf p}_l,{\bf p}_{\bar{\nu}}$) in the rest frame of the $B$-meson).
This is the set of variables initially introduced by Cabibbo and Maksymowicz
 \cite{cabibbo} in the analysis of $K_{l4}$ decays.
All the angles above are explicitly defined in Fig.~2.
Then, the remaining invariants are
\be
&&P\cdot N=\frac{1}{2s_L}\Big[m_l^2(m_B^2-s_M-s_L)
       +\cos\theta_l(s_L-m_l^2)\lambda^{1/2}(\MB^2,s_M,s_L)\Big],\nonumber\\
&&Q\cdot L=\frac{1}{2s_M}\Big[(\MD^2-m_\pi^2)(\MB^2-s_M-s_L)
         +\cos\theta\lambda^{1/2}(s_M,\MD^2,m_\pi^2)
          \lambda^{1/2}(\MB^2,s_M,s_L)\Big],\nonumber\\
&&Q\cdot N=\frac{m_l^2}{2s_Ms_L}\Big[(\MD^2-m_\pi^2)(\MB^2-s_M-s_L)
    +\cos\theta\lambda^{1/2}(\MB^2,s_M,s_L)\lambda^{1/2}(s_M,\MD^2,m_\pi^2)\Big]\nonumber\\
  &&\hskip 1.5cm +\frac{\MD^2-m_\pi^2}{2s_Ms_L}\cos\theta_l\lambda^{1/2}(\MB^2,s_M,s_L)
           \lambda^{1/2}(s_L,m_l^2,0)\nonumber\\
        &&\hskip 1.5cm+\frac{(\MB^2-s_M-s_L)}{2s_Ms_L}\cos\theta_l\cos\theta
        \lambda^{1/2}(s_M,\MD^2,m_\pi^2)\lambda^{1/2}(s_L,m_l^2,0)\nonumber\\
       &&\hskip 1.5cm -\frac{1}{\sqrt{s_Ms_L}}\cos\phi\sin\theta_l\sin\theta
         \lambda^{1/2}(s_M,\MD^2,m_\pi^2)\lambda^{1/2}(s_L,m_l^2,0),\nonumber\\
&&\e_{\mu\nu\rho\sigma}P^\mu Q^\nu L^\rho N^\sigma\nonumber\\
     &&\hskip 0.5cm =-\frac{1}{2\sqrt{s_Ls_M}}\lambda^{1/2}(\MB^2,s_M,s_L)
  \lambda^{1/2}(s_M,\MD^2,m_\pi^2)\lambda^{1/2}(s_L,m_l^2,0)\sin\phi\sin\theta_l\sin\theta ,
\label{invariant}
\ee
where $\lambda(x,y,z)=x^2+y^2+z^2-2xy-2yz-2xz$.
In particular, the amplitudes are functions of three invariants,
$\o \equiv v\cdot v'$, $p\cdot v$ and $p\cdot v'$, which are given by
\be
&&v\cdot v'=\frac{1}{4\MB\MD s_M}\Big[(s_M+\MD^2-m_\pi^2)(\MB^2+s_M-s_L)\nonumber\\ 
&&\hskip 3.8cm +\lambda^{1/2}(\MB^2,s_M,s_L)\lambda^{1/2}(s_M,\MD^2,m_\pi^2)\cos\theta\Big],\nonumber\\
&&p\cdot v=\frac{1}{4\MB s_M}\Big[(s_M+m_\pi^2-\MD^2)(\MB^2+s_M-s_L)\nonumber\\ 
&&\hskip 3.0cm -\lambda^{1/2}(\MB^2,s_M,s_L)\lambda^{1/2}(s_M,\MD^2,m_\pi^2)\cos\theta\Big],\nonumber\\
&&p\cdot v'=\frac{1}{2\MD}\Big(s_M-\MD^2-m_\pi^2\Big).
\ee

Finally, the decay rate is proportional to
\be
|T|^2=\tilde{\kappa}^2 L^{\mu\nu}\Omega_\mu\Omega_\nu^*,
\ee
where $\tilde{\kappa}=\kappa |1+\chi|$, and the leptonic tensor $L_{\mu\nu}$ 
is given by
\be
L_{\mu\nu}=4(L_\mu L_\nu-N_\mu N_\nu -(s_L-m_l^2)g_{\mu\nu}
     -i\e_{\mu\nu\rho\sigma}L^\rho N^\sigma),
\ee
retaining lepton mass $m_l$.
The explicit result is
\be
&&|T|^2/\tilde{\kappa}^2=4|F|^2[(P\cdot L)^2-(s_L-m_l^2)s_M-(P\cdot N)^2]
              +4|G|^2[(Q\cdot L)^2-Q^2(s_L-m_l^2)\nonumber\\
              &&\hskip 1.5cm -(Q\cdot N)^2]
  +|H|^2[(s_L-m_l^2)(2P\cdot LP\cdot QQ\cdot L-(P\cdot L)^2Q^2
    -(P\cdot Q)^2s_L\nonumber\\
   &&\hskip 1.5cm -(Q\cdot L)^2s_M+Q^2s_M s_L)
    -(\e_{\mu\nu\rho\sigma}L^\mu N^\nu P^\rho Q^\sigma)^2]
     +8\Re(FG^*)[P\cdot LQ\cdot L\nonumber\\
  &&\hskip 1.5cm -P\cdot NQ\cdot N-(s_L-m_l^2)
      P\cdot Q]+4\Re(FH^*)[s_LP\cdot QP\cdot N+(P\cdot L)^2Q\cdot N\nonumber\\
  &&\hskip 1.5cm -P\cdot LQ\cdot L
   P\cdot N-s_M s_L Q\cdot N
   -m_l^2(P\cdot QP\cdot L-s_MQ\cdot L)]\nonumber\\
  &&\hskip 1.5cm 
  +4\Re(GH^*)[-P\cdot N(Q\cdot L)^2+P\cdot LQ\cdot LQ\cdot N+Q^2s_LP\cdot N
    -s_Lp\cdot QQ\cdot N\nonumber\\
    &&\hskip 1.5cm -m_l^2(Q^2P\cdot L-P\cdot QQ\cdot L)]
    +4m_l^2[|R|^2(s_L-m_l^2)+2\Re(F^*R)(P\cdot L-P\cdot N)\nonumber\\
  &&\hskip 1.5cm +2\Re(G^*R)(Q\cdot L-Q\cdot N)
    -\Im(RH^*)\e_{\mu\nu\rho\sigma}P^\mu Q^\nu L^\rho N^\sigma]\nonumber\\
  &&\hskip 1.5cm +4\Im(2FG^*+F^*HP\cdot N+G^*HQ\cdot N)\e_{\mu\nu\rho\sigma}
    P^\mu Q^\nu L^\rho N^\sigma\;.
\label{amp}
\ee

The differential partial width of interest can be expressed as
\be
d\Gamma_{B_{l4}}=\frac{N_\pi}{2\MB}J(s_M,s_L)|T|^2 d \Phi_4,
\ee
where the 4 body phase space $d \Phi_4$ is
\be
d \Phi_4 \equiv  ds_M \cdot ds_L \cdot d\cos\theta \cdot d\cos\theta_l \cdot d\phi,
\ee
and
\be
N_\pi = \left\{\begin{array}{cc}
            2 & {\rm for\; charged\; pions,}\\
            1 & {\rm for\; neutral\; pions,}
            \end{array}\right.
\ee
and the Jacobian $J(x,y)$ is
\be
J(x,y)=\frac{1}{2^{14}\pi^6xy\MB^2}\lambda^{1/2}(\MB^2,x,y)
      \lambda^{1/2}(x,\MD^2,m_\pi^2)\lambda^{1/2}(y,m_l^2,0).
\ee
Kinematically allowed regions of the variables are
\be
&&(m_D+m_\pi)^2\;<\;s_M\;<\;(m_B-m_l)^2,\nonumber\\
&&m_l^2\;<\;s_L\;<\;(m_B-\sqrt{s_M})^2,\nonumber\\
&&-1\;<\;\cos\theta,\;\cos\theta_l\;<\;1,\nonumber\\
&&0\;<\;\phi\;<\;2\pi .
\ee

Since the initial $B^-$ system is not CP self-conjugate, any genuine
CP-odd observable can be constructed only by considering both the $B^-_{l4}$
decay and its charge-conjugated $B^+_{l4}$ decay, and by identifying the CP
relations of their kinematic distributions.
Before constructing possible CP-odd asymmetries explicitly, we calculate
the transition probability for the charge-conjugated process
$B^+\to \overline{D^0} \pi^0 l^+ \nu_l$ or $B^+\to D^- \pi^+ l^+ \nu_l$. 
Following previous explicit expressions,
the transition probability $|\overline{T}|^2$ for the $B^+$ decay in the same 
reference frame as in the $B^-$ decay is given by simple modification of the 
transition probability $|T|^2$ of the $B^-$ decay:
\be
|\overline{T}|^2=|T|^2\left\{\begin{array}{l}
   {\rm (i)\; change\; signs\; in\; front\; of\; the\; terms\; proportional\; 
    to\; imaginary\; part}\\
   {\rm (ii)\;}\zeta\to\zeta^*\;.\end{array}\right.
\label{ampbar}
\ee
It is easy to see that if the parameter $\zeta$ is real, the transition
probability (\ref{amp}) for the $B^-$ decay and (\ref{ampbar}) for
the $B^+$ decay satisfy CP relation:
\be
|T|^2(\theta,\; \theta_l,\; \phi)=|\overline{T}|^2(\theta,\; \theta_l,\; -\phi).
\label{cprel}
\ee 
Note that all the imaginary parts are being multiplied by
the quantity $\e_{\mu\nu\rho\sigma}P^\mu Q^\nu L^\rho N^\sigma$ 
which is proportional
to $\sin\phi$, as can be seen in Eq.~(\ref{invariant}).
And then, 
$d\Gamma/d\Phi_4$ can be decomposed into a CP-even part ${\cal S}$ and
a CP-odd part ${\cal D}$:
\be
\frac{d\Gamma}{d\Phi_4}=\frac{1}{2}({\cal S}+{\cal D}).
\ee
The CP-even part ${\cal S}$ and the CP-odd part ${\cal D}$ can be easily
identified by making use of the CP relation (\ref{cprel}) between the $B^-$ and 
$B^+$ decay probabilities and they are expressed as
\be
{\cal S} =\frac{d(\Gamma+\overline{\Gamma})}{d\Phi_4},\;
~~~{\rm and}~~~{\cal D} =\frac{d(\Gamma-\overline{\Gamma})}{d\Phi_4},
\ee
where we have used the same kinematic variables $\{s_M,s_L,\theta,\theta_l\}$
for the $d\overline{\Gamma}/d\Phi_4$ except for the replacement of ${\phi}$ 
by $-\phi$, as shown in Eq. (\ref{cprel}).
Here $\Gamma$ and $\overline{\Gamma}$ are the decay rates for $B^-$ and $B^+$, 
respectively. The CP-even ${\cal S}$ term and the CP-odd ${\cal D}$ term can be obtained from
$B^\mp$ decay probabilities and their explicit form is listed in Appendix B.
Note that the CP-odd term is proportional to the imaginary part 
of the parameter $\zeta$ in Eq.~(\ref{zeta}) and lepton mass $m_l$ 
[Note also that ${\cal D}$ in
Eq.~(\ref{delta}) looks like proportional to $m_l^2$, 
but the quantity $A_4$ has $1/m_l$ factor
in it].
Therefore, there exists no CP violation in \bl decays within the SM since the SM
corresponds to the case with $\zeta =0$.

\section{Asymmetries}

\noindent
An easily-constructed CP-odd asymmetry is the rate asymmetry
\be
A\equiv \frac{\Gamma-\overline{\Gamma}}{\Gamma+\overline{\Gamma}},
\ee
which has been used as a 
probe of CP violation in Higgs and top quark sectors  \cite{rateasym}.  
Here $\Gamma$ and $\overline{\Gamma}$ are the decay rates for $B^-$ and $B^+$, 
respectively.
The statistical significance of the asymmetry can then be computed as
\be
N_{SD}=\frac{N_- -N_+}{\sqrt{N_-+N_+}} =\frac{N_- -N_+}{\sqrt{N \cdot Br}},
\ee
where $N_{SD}$ is the number of standard deviations,
$N_\pm$ is the number of events predicted in $B_{l4}$ decay for $B^\pm$
meson, $N$ is the number of $B$ mesons produced,
and $Br$ is the branching fraction of the relevant $B$ decay mode. 
For a realistic detection efficiency
$\e$, we have only to rescale the number of events by this parameter,
$N_-+N_+\to\e (N_-+N_+)$. Taking $N_{SD}=1$, we obtain 
the number $N_B$ of the B mesons needed to observe CP violation at $1$-$\sigma$ level:
\be
N_B=\frac{1}{Br\cdot A^2}.
\ee

Next, we consider the so-called optimal observable.
An appropriate real weight function $w(s_M,s_L;\theta,\theta_l,\phi)$
is usually employed to separate the CP-odd ${\cal D}$ contribution and to enhance
its analysis power for the CP-odd parameter ${\rm Im}(\zeta)$ through 
the CP-odd quantity:
\begin{eqnarray}
\langle w{\cal D}\rangle\equiv\int\left[w{\cal D}\right] d\Phi_4,
\end{eqnarray}
of which the analysis power is determined by the parameter 
\begin{eqnarray}
\varepsilon
   =\frac{\langle w{\cal D}\rangle}{\sqrt{\langle{\cal S}\rangle
          \langle w^2{\cal S}\rangle}}\;.
\label{Significance}
\end{eqnarray}
For the analysis power $\varepsilon$, the number $N_B$ of the $B$ mesons 
needed to observe CP violation at 1-$\sigma$ level is
\begin{eqnarray}
N_B=\frac{1}{Br\cdot\varepsilon^2}\;.
\label{eq:number}
\end{eqnarray}
Certainly, it is desirable to find the optimal weight function
with the largest analysis power. It is known  \cite{Optimal} that 
when the CP-odd contribution to the total rate is relatively small, 
the optimal weight function  is approximately given as
\begin{eqnarray}
w_{\rm opt}(s_M,s_L;\theta,\theta_l,\phi)=
 \frac{{\cal D}}{{\cal S}}~~~\Rightarrow~~~
 \varepsilon_{\rm opt}=\sqrt{\frac{\langle\frac{{\cal D}^2}{{\cal S}}\rangle}
 {\langle{\cal S}\rangle}}.
\end{eqnarray}
We adopt this optimal weight function in the following numerical analyses.

\section{Numerical Results and Conclusions}

\noindent
Let us first consider the decay to heavy lepton, $\tau$, since CP-odd asymmetry is 
proportional to $m_l$, and heavy lepton may be more susceptible to effects
of new physics.
The contributions of each multiplet to the $B_{\tau 4}$ decays are shown
in Figs.~3(a) and (b), for
$B^-\to D^+\pi^-\tau\bar{\nu}_\tau$ and
$B^-\to D^0\pi^0\tau\bar{\nu}_\tau$, respectively.
We found that the contributions from $(0^+,1^+)$ and $(0^-,1^-)'$ 
multiplets are comparable
to the lowest one $(0^-,1^-)$, and 
the contribution of the D-wave $(1^-,2^-)$ multiplet 
is very small. However, 
the latter may give non-negligible contribution through the interference
with the other dominant parts, so we retain it.

We restrict ourselves to the soft-pion limit by considering only the region
$s_M\le 6.5~ {\rm GeV}^2$, which is about one half of the maximum value.
This restriction corresponds to the pion momentum less than about $0.6$ GeV.
The decay rates we obtain (including all of the resonances we have discussed)
are $1.79\times 10^{-16}~{\rm GeV}$ for $B^-\to D^+\pi^-\tau\bar{\nu}_\tau$
and $5.56\times 10^{-15}~{\rm GeV}$ for $B^-\to D^0\pi^0\tau\bar{\nu}_\tau$,
where we use $\Gamma_{D^{*0}}=60\;{\rm keV}$ 
for the total width of the $D^{*0}$ \cite{goity}.
These correspond to branching fractions of 
$0.045\%$ and $1.37\%$, respectively,
which were obtained by using the recently published lifetime of the 
charged $B$-meson ($\tau_{B^\pm}=1.65\times 10^{-12} s$) \cite{pdg}.
We notice that, although $B^-\to D^+\pi^-\tau\bar{\nu}_\tau$
decay amplitude is larger than that of $B^-\to D^0\pi^0\tau\bar{\nu}_\tau$
by factor $\sqrt{2}$ from isospin relation, for the actual decay rate 
the latter mode has much larger decay rate.
This is so because the most dominant resonance $D^{0*}$  cannot
decay into $D^+\pi^-$ on its mass-shell \cite{pdg}.

In Table 1, we show the results of $B_{\tau 4}$ decays 
for the two CP-violating asymmetries:
the rate asymmetry $A$ and the optimal asymmetry $\varepsilon_{\rm opt}$.
We estimated the number of $B$ meson pairs, $N_B$, needed for detection at 
$1\sigma$ level for CP-violating values $|\Im(\zeta)|=0.5$ and $2.0$.
These values are chosen from our rough estimates of current experimental bounds 
in the multi-Higgs-doublet model and scalar leptoquark models.
Thorough investigation on specific models can be found in Ref.~\cite{bcp}.
As we can expect, the optimal observable gives much better result than
the simple rate asymmetry. We also found that although the decay rate of neutral pion
mode is larger than that of the charged pion mode, 
the latter case gives better detection results 
because of large CP-violating effects in charged pion decay mode.
Since one expect about $10^8$ orders of $B$-meson pairs produced yearly in the 
asymmetric $B$ factories, 
one could probe the CP-violation effect by using the optimal asymmetry observable.

We also estimated CP-violation effects in the \bl decays with light leptons.
The results for $B_{\mu 4}$ decays are shown in Table 2.
We found that relative sizes of CP-violating asymmetry, which is proportional
to the lepton mass, are smaller in $B_{\mu 4}$ decays than in $B_{\tau 4}$ decays.
However, due to the larger branching fractions, at the same value of $\Im(\zeta)$
one needs in the $B_{\mu 4}$ case the number of $B$-meson pairs of the same
order of magnitude as in the $B_{\tau 4}$ case to probe the CP-violating effect.
For $B_{e 4}$ decays, however, electron mass is so small 
that we find that $\sim$$10^{10}$ 
$B$-meson pairs are needed for the same input $\Im(\zeta)$ values, 
even when using the optimal observable.
Thus one may conclude that both $B_{\tau 4}$ and $B_{\mu 4}$ decays could serve as
equally good probes of CP-violation effects. However, 
in some models such as multi-Higgs-doublet models, 
the scalar coupling, {\it i.e.\/} effectively $\zeta$ itself, is
proportional to the lepton mass, 
and therefore the CP violation in the $B_{\mu 4}$ decays is
highly suppressed compared with the $B_{\tau 4}$ decays.
Therefore, $B_{\tau 4}$ decays could serve as a much better probe of CP-violating
effects in such models.

As mentioned earlier, the pure contribution of the $D$-wave $(1^-,2^-)$ multiplet
is very small (cf.~Fig.~3), but we included that multiplet
since its contribution could be rather
sizable through interference with other dominant ones.
However, we compared the result obtained by dropping off the $(1^-,2^-)$ multiplet
with the previous results including all the multiplets,
and found that the interference effect is also very small.
So one could safely neglect the $(1^-,2^-)$ multiplet for $B_{l4}$ decays.

We here included various higher excited states other than just the ground state
$D^*$ vector mesons as intermediate states in the decay processes.
In order to estimate effects of these higher excited states on CP violation effects,
we show in Table 3 the results with only the ground state included 
as an intermediate state.
It can be easily seen, by comparing with the values in Table 1(b),
that those higher excited states have highly amplified the effect of CP violation.

In conclusion, we investigated
CP violation from physics beyond the Standard Model through semileptonic $B_{l4}$
decays: $B\to D\pi l\bar{\nu}_l$. In the decay process, we included various excited states 
as intermediate states decaying to the 
final hadrons. The CP violation is implemented 
in a model independent way, 
in which we extend the leptonic current
by including complex couplings  of the scalar sector and those of the vector sector
in extensions of the Standard Model.
We calculated the CP-odd rate asymmetry and the optimal asymmetry, and
found that the optimal asymmetry is sizable and
can be detected at $1\sigma$ level 
with about $10^6$-$10^7$ $B$-meson pairs, 
for some reference values of new physics effect.
Since $\sim$$10^8$ $B$-meson pairs are expected 
to be produced yearly at the forthcoming asymmetric $B$ factories,
one could investigate CP-violation effects by using our formalism.
\\

\section*{Acknowledgments}

\noindent
We thank H. Y. Cheng and G. Cveti\v{c} for careful reading of the manuscript and 
their valuable comments.
CSK wishes to thank the Korea Institute for Advanced Study for warm
hospitality. The work of CSK was supported 
in part by Non-Directed-Research-Fund of 1997, KRF,
in part by the CTP, Seoul National University, 
in part by the BSRI Program, Ministry of Education, Project No. BSRI-98-2425,
in part by the KOSEF-DFG large collaboration project, 
Project No. 96-0702-01-01-2.
JL wishes to acknowledge the financial support of 
Korean Research Foundation made in the program of 1997.
The work of WN was supported by the BSRI Program, 
Ministry of Education, Project No. BSRI-98-2425.

\newpage

\appendix

\begin{appendix}

\section{Form factors}

We give the form factors needed in Eq.~(\ref{Rho2}).
If we write
\be
\Omega_\mu=ih\MB\MD\e_{\mu\nu\rho\sigma}v^\nu v'^\rho p^\sigma+A_1 p_\mu
+A_2\MB v_\mu
           +A_3\MD v'_\mu +\zeta A_4 L_\mu\;,
\ee
then, the form factors in Eq.~(\ref{Rho2}) are expressed as
\be
H&=&h,\nonumber\\
F&=&A_2+\frac{1}{2}(A_1+A_3),\nonumber\\
G&=&\frac{1}{2}(A_3-A_1),\nonumber\\
R&=&A_2+\zeta A_4,
\ee
where
\be
h&=&\frac{g\xi(\o)}{f_\pi \MB\MD}\Big(\Delta(\Bst)+\Delta(\Dst)\Big)
    +\frac{\alpha_2\rho_2(\o)}{3f_\pi \MB\MD}(\o-1)\Big(\Delta(\Bone)
    +\Delta(\Done)\Big)
     \nonumber\\
  &&+\frac{\alpha_3\xi^{(1)}(\o)}{f_\pi \MB\MD}\Big(\Delta(\Bp)+\Delta(\Dp)\Big),\\
A_1&=&\frac{g\xi(\o)}{f_\pi }(1+\o)\Big(\Delta(\Bst)+\Delta(\Dst)\Big)
      +\frac{\alpha_2\rho_2(\o)}{3f_\pi }(\o^2-1)\Big(\Delta(\Bone)
      +\Delta(\Done)\Big)
       \nonumber\\
    &&+\frac{\alpha_3\xi^{(1)}(\o)}{f_\pi }(1+\o)\Big(\Delta(\Bp)+\Delta(\Dp)\Big),\\
A_2&=&-\frac{g\xi(\o)}{f_\pi \MB}p\cdot(v+v')\Delta(\Bst)
    -\frac{\alpha_1\rho_1(\o)}{f_\pi \MB}
     \Big(p\cdot v\Delta(\Bnot)-p\cdot v'\Delta(\Dnot)\Big)
     \nonumber\\
   &&-\frac{\alpha_2\rho_2(\o)}{f_\pi \MB}\bigg\{\Delta(\Bone)
     \Big[\frac{1}{3}(\o p\cdot v'-p\cdot v)+\frac{2}{3}(p\cdot v'-\o p\cdot v)\Big]
     -\Delta(\Done)(p\cdot v-\o p\cdot v')\bigg\}\nonumber\\
   &&-\frac{\alpha_3\xi^{(1)}(\o)}{f_\pi \MB}p\cdot(v+v')\Delta(\Bst),\\
A_3&=&-\frac{g\xi(\o)}{f_\pi \MD}p\cdot(v+v')\Delta(\Dst)
    -\frac{\alpha_1\rho_1(\o)}{f_\pi \MD}
     \Big(p\cdot v'\Delta(\Dnot)-p\cdot v\Delta(\Bnot)\Big)
     \nonumber\\
   &&-\frac{\alpha_2\rho_2(\o)}{f_\pi \MD}\bigg\{\Delta(\Done)
     \Big[\frac{1}{3}(\o p\cdot v-p\cdot v')+\frac{2}{3}(p\cdot v-\o p\cdot v')\Big]
     -\Delta(\Bone)(p\cdot v'-\o p\cdot v)\bigg\}\nonumber\\
   &&-\frac{\alpha_3\xi^{(1)}(\o)}{f_\pi \MD}p\cdot(v+v')\Delta(\Dst),\\
A_4&=&\frac{g\xi(\o)}{f_\pi m_l}\Big[-\Delta(\Bst)(p\cdot v'-\o p\cdot v)
     +\Delta(\Dst)(p\cdot v-\o p\cdot v')\Big]\nonumber\\
    &&+\frac{\alpha_1\rho_1(\o)}{f_\pi m_l}(1-\o)\Big[-p\cdot v\Delta(\Bnot)
      +p\cdot v'\Delta(\Dnot)\Big]\nonumber\\
    &&+\frac{2\alpha_2\rho_2(\o)}{3f_\pi m_l}
    (1-\o)\Big[-\Delta(\Bone)(p\cdot v'-\o p\cdot v)
     +\Delta(\Done)(p\cdot v-\o p\cdot v')\Big]\nonumber\\
  &&+\frac{\alpha_3\xi^{(1)}(\o)}{f_\pi m_l}\Big[-\Delta(\Bp)(p\cdot v'-\o p\cdot v)
     +\Delta(\Dp)(p\cdot v-\o p\cdot v')\Big].
\ee
\section{CP-even and CP-odd quantities}
The CP-even quantity ${\cal S}$ is
\be
&&{\cal S}=2C(s_M,s_L)\cdot (R{\rm -indep.\; terms\; of\;} |T|^2)\nonumber\\
  &&\hskip 0.8cm +8C(s_M,s_L)m_l^2\Big[ 2(s_L-m_l^2)\Re(\zeta)\Re(A_2A_4^*)
   +2(P\cdot L-P\cdot N)\{\Re(F^*A_2)\nonumber\\
  &&\hskip 0.8cm +\Re(\zeta)\Re(F^*A_4)\}
    +2(Q\cdot L-Q\cdot N)\{\Re(G^*A_2)+\Re(\zeta)\Re(G^*A_4)\}\nonumber\\
  &&\hskip 0.8cm -\e_{\mu\nu\rho\sigma}P^\mu Q^\nu L^\rho N^\sigma
        \{\Re(H^*A_2)+\Re(\zeta)\Re(H^*A_4)\}\Big],
\ee
and the CP-odd quantity ${\cal D}$ is
\be
&&{\cal D}=8C(s_M,s_L)m_l^2\Im(\zeta)\Big[2(s_L-m_l^2)\Im(A_2A_4^*)
       -2(P\cdot L-P\cdot N)\Im(F^*A_4)\nonumber\\
      &&\hskip 0.8cm -2(Q\cdot L-Q\cdot N)\Im(G^*A_4)
       +\e_{\mu\nu\rho\sigma}P^\mu Q^\nu L^\rho N^\sigma\Im(H^*A_4)\Big], 
\label{delta}
\ee
where the overall function $C(s_M,s_L)$ is given by
\be
C(s_M,s_L)=\kappa^2|1+\chi|^2\frac{N_\pi}{2m_B}J(s_M,s_L).
\ee
\end{appendix}
%
%

\begin{table}
{Table~1}. {The CP-violating rate asymmetry A and the optimal asymmetry
$\varepsilon_{\rm opt}$, determined in the soft pion limit, and the number
of charged $B$ meson pairs, $N_B$, needed for detection at $1\sigma$ level,
at reference values (a) $\Im(\zeta)=0.5$ and (b) $\Im(\zeta)=2.0$,
for the $B_{\tau 4}$ decays.}\par
\begin{tabular}{c|cc|cc}
(a) Modes  
& \multicolumn{2}{c|}{$B^-\to D^+\pi^-\tau\bar{\nu}_\tau$}
& \multicolumn{2}{c}{$B^-\to D^0\pi^0\tau\bar{\nu}_\tau$}\\
\hline\hline
Asymmetry & Size(\%) & $N_B$ & Size(\%) & $N_B$ \\
\hline
A & $0.08$ & $3.31\times 10^{9}$ & $0.002$ & $3.22\times 10^{11}$\\
$\varepsilon_{\rm opt}$ & $1.13$ & $1.77\times 10^7$ & $0.15$ & $3.33\times 10^7$\\
\hline\hline
(b) Modes  
& \multicolumn{2}{c|}{$B^-\to D^+\pi^-\tau\bar{\nu}_\tau$}
& \multicolumn{2}{c}{$B^-\to D^0\pi^0\tau\bar{\nu}_\tau$}\\
\hline\hline
Asymmetry & Size(\%) & $N_B$ & Size(\%) & $N_B$ \\
\hline
A & $0.33$ & $2.07\times 10^{8}$ & $0.006$ & $2.01\times 10^{10}$\\
$\varepsilon_{\rm opt}$ & $4.52$ & $1.11\times 10^6$ & $0.59$ & $2.08\times 10^6$\\
\end{tabular}
\end{table}
\begin{table}
{Table~2}. {The CP-violating rate asymmetry A and the optimal asymmetry
$\varepsilon_{\rm opt}$, determined in the soft pion limit, and the number
of charged $B$-meson pairs, $N_B$, needed for detection at $1\sigma$ level,
at reference values (a) $\Im(\zeta)=0.5$ and (b) $\Im(\zeta)=2.0$,
for the $B_{\mu 4}$ decays.}\par
\begin{tabular}{c|cc|cc}
(a) Modes  
& \multicolumn{2}{c|}{$B^-\to D^+\pi^-\mu\bar{\nu}_\mu$}
& \multicolumn{2}{c}{$B^-\to D^0\pi^0\mu\bar{\nu}_\mu$}\\
\hline\hline
Asymmetry & Size(\%) & $N_B$ & Size(\%) & $N_B$ \\
\hline
A & $0.014$ & $7.27\times 10^{9}$ & $0.0005$ & $3.57\times 10^{11}$\\
$\varepsilon_{\rm opt}$ & $0.24$ & $2.35\times 10^7$ & $0.045$ & $4.53\times 10^7$\\
\hline\hline
(b) Modes  
& \multicolumn{2}{c|}{$B^-\to D^+\pi^-\mu\bar{\nu}_\mu$}
& \multicolumn{2}{c}{$B^-\to D^0\pi^0\mu\bar{\nu}_\mu$}\\
\hline\hline
Asymmetry & Size(\%) & $N_B$ & Size(\%) & $N_B$ \\
\hline
A & $0.054$ & $4.54\times 10^{8}$ & $0.002$ & $1.83\times 10^{10}$\\
$\varepsilon_{\rm opt}$ & $0.95$ & $1.47\times 10^6$ & $0.2$ & $2.83\times 10^6$\\
\end{tabular}
\end{table}
\begin{table}
{Table~3}. {The CP-violating rate asymmetry A and the optimal asymmetry
$\varepsilon_{\rm opt}$, determined in the soft pion limit, and the number
of charged $B$-meson pairs, $N_B$, needed for detection at $1\sigma$ level,
at the reference value $\Im(\zeta)=2.0$,
for the $B_{\tau 4}$ decays, including only the ground state $(0^-,1^-)$ multiplet
as an intermediate state.}\par
\begin{tabular}{c|cc|cc}
Modes  
&\multicolumn{2}{c|}{$B^-\to D^+\pi^-\tau\bar{\nu}_\tau$}
&\multicolumn{2}{c}{$B^-\to D^0\pi^0\tau\bar{\nu}_\tau$}\\
\hline\hline
 Asymmetry & Size(\%) & $N_B$ & Size(\%) & $N_B$ \\
\hline
 A & $0.0004$ & $2.20\times 10^{14}$ & $0.0002$ & $1.75\times 10^{13}$ \\
 $\varepsilon_{\rm opt}$ & $0.007$ & $6.06\times 10^{11}$ & $0.01$ & $5.39\times 10^9$ \\
\end{tabular}
\end{table}

%
\newpage
\parindent=0 cm
%
%
%
%
\begin{figure}[h]
\vspace*{-2.0cm}
\hbox to\textwidth{\hss\epsfig{file=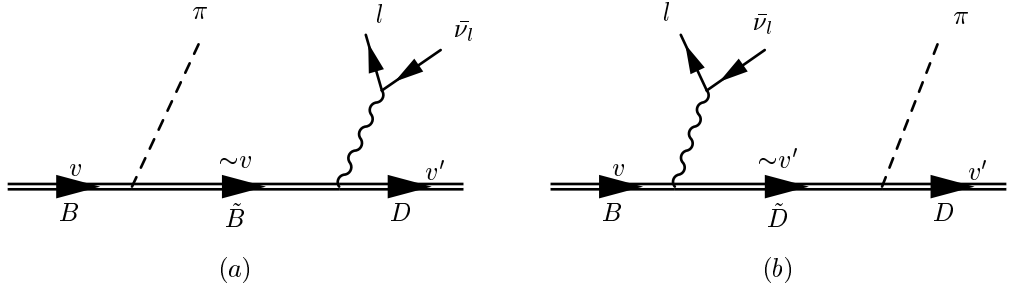,height=30cm}\hss}
\vspace*{-21.5cm}
\caption{Feynman diagrams for $B_{l4}$ decays.}
\label{fig:diagram}
\end{figure}

\vspace*{3.0cm}

\begin{figure}[h]
\hbox to\textwidth{\hss\epsfig{file=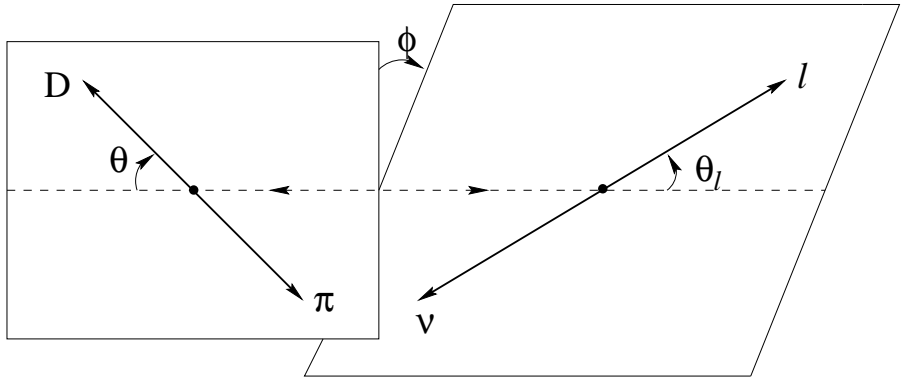,height=5cm}\hss}
\vspace*{1.0cm}
\caption{Decay planes for the kinematic variables}
\label{fig:plane}
\end{figure}

\begin{figure}[h]
\vspace*{-4cm}
\centerline{\epsfig{figure=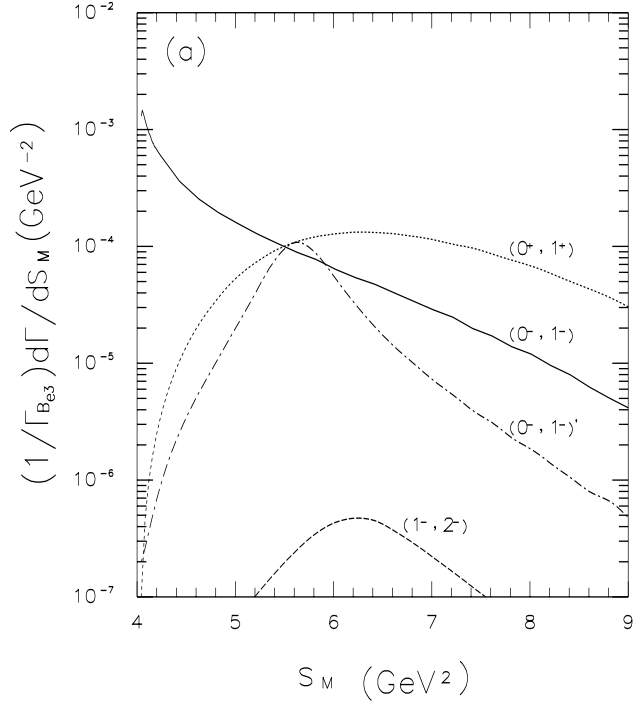,height=19cm,width=15cm,angle=0}}
\vspace*{-9cm}
\centerline{\epsfig{figure=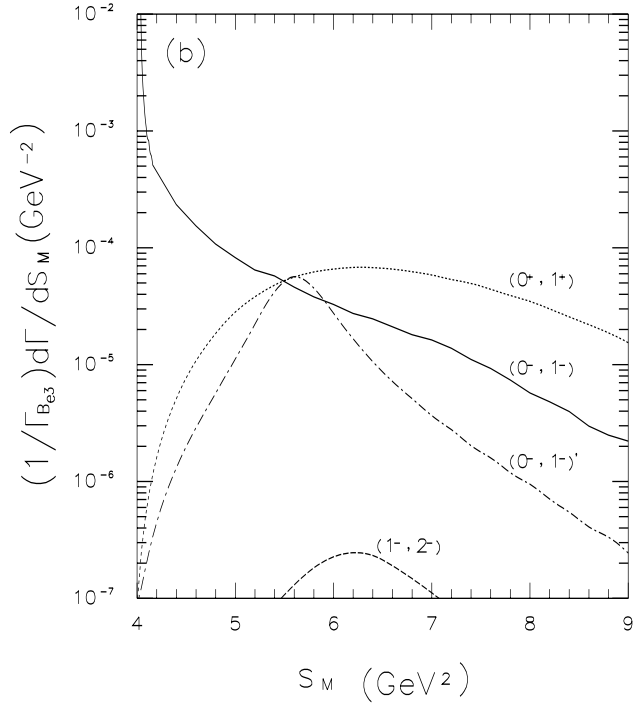,height=19cm,width=15cm,angle=0}}
\vspace*{-5cm}
\caption{$\frac{1}{\Gamma(B\to De\nu)}\frac{d\Gamma_{B\tau 4}}{ds_M}$
    as a function of the invariant mass of $D\pi$, $s_M$: 
(a) for $B^-\to D^+\pi^-\tau\nu$ and
(b) for $B^-\to D^0\pi^0\tau\nu$.
The curves correspond to the individual resonance doublets, as indicated
in the figures.}
\end{figure}

\end{document}